\pdfoutput=1

\documentclass{optica-article}
\journal{opticajournal}
\articletype{Research Article}
\usepackage{float}

\begin{document}

\title{Coherent EUV scatterometry of 2D periodic structure profiles with mathematically optimal experimental design}

\author{Clay Klein,\authormark{1,*} Nicholas W. Jenkins,\authormark{1} Yunzhe Shao,\authormark{1} Yunhao Li,\authormark{1} Seungbeom Park,\authormark{2} Wookrae Kim,\authormark{3} Henry C. Kapteyn,\authormark{1,4} and Margaret M. Murnane\authormark{1}}

\address{
\authormark{1}Department of Physics, JILA \& STROBE NSF Science \& Technology Center, University of Colorado \& NIST, Boulder, Colorado, USA\\
\authormark{2}Core Technology R\&D Team, Mechatronics Research, Samsung Electronics Co., Ltd., Hwasung, Republic of Korea\\
\authormark{3}Advanced Process Development Team 4, Semiconductor R\&D Center, Samsung Electronics Co., Ltd., Hwasung, Republic of Korea\\
\authormark{4}KMLabs, Inc., 4775 Walnut Street, \#102, Boulder, Colorado 80301, USA
}

\email{\authormark{*}clay.klein@colorado.edu}

\begin{abstract*} 
Extreme ultraviolet (EUV) scatterometry is an increasingly important metrology that can measure critical parameters of periodic nanostructured materials in a fast, accurate, and repeatable manner and with high sensitivity to nanoscale structure and material composition. Because of this, EUV scatterometry could support manufacturing of semiconductor devices or polymer metamaterials, addressing the limitations of traditional imaging methods such as resolution and field of view, sample damage, throughput, or low sensitivity. Here we use EUV scatterometry to measure the profile of an industrially relevant 2D periodic interconnect structure, using $\lambda=29$ nm light from a table-top high harmonic generation source. We show that EUV scatterometry is sensitive to out-of-plane features with single-nanometer sensitivity. Furthermore, we also apply a methodology based on the Fisher information matrix to optimize experimental design parameters, such as incidence angles and wavelength, to show how measurement sensitivity can be maximized. This methodology reveals the strong dependence of measurement sensitivity on both incidence angle and wavelength — even in a simple two-parameter case. Through a simultaneous optimization of incidence angles and wavelength, we determine that the most sensitive measurement of the quantities of interest can be made at a wavelength of $\sim$14 nm. In the future, by reducing sample contamination due to sample preparation, deep sub-nanometer sensitivity to axial profiles and 2D structures will be possible. Our results are an important step in guiding EUV scatterometry towards increased accuracy and throughput with \textit{a priori} computations and by leveraging new experimental capabilities.
\end{abstract*}

\section{Introduction}
The characterization of periodic nanostructured materials is a critical capability for understanding fundamental material properties as well as for the fabrication of semiconductor devices, optical metamaterials, and lithographic masks \cite{esashi2023tabletop,wang2023high,klein2024extreme,jenkins2024euv,soltwisch2017reconstructing,porter2023soft,raymond2005overview,gardner2017subwavelength,nelson2024tabletop,knobloch2022structural,nakasuji2012development,sabbagh2023optical,sherwin2024sub,shen2024spectral,shen2024euv,ku2016euv}. Coherent extreme ultraviolet (EUV) light generated through high harmonic generation (HHG) of ultrafast laser pulses is a source particularly well suited for such measurements due to their broad spectral coverage $\sim$10-100 nm, high coherence \cite{rundquist1998phase,bartels2002generation}, and high stability. EUV HHG scatterometry and coherent diffractive imaging (CDI) are sensitive to nanoscale topographic features and buried layers with good elemental and chemical specificity due to a large phase contrast in reflection for many common scientifically interesting and industrially useful materials \cite{esashi2023tabletop,wang2023high,klein2024extreme,shanblatt2016quantitative,jansen2019broadband,miao2015beyond,loetgering2022advances,tanksalvala2021nondestructive,doering2012euv,eschen2024structured}.

Scatterometry is a technique that analyzes the spatial distribution of scattered light from a sample. When EUV light is used as the probe, the technique can yield quantitative results when combined with appropriate scattering models. Furthermore, refractive indices of many materials can be computed a priori through tabulated scattering factors \cite{henke1993x}, providing the flexibility to analyze samples without extensive prior characterization as is often required when using visible wavelengths where the index of refraction depends strongly on the chemical binding or band structure of the material.

EUV scatterometry is thus a powerful tool that is gaining increased interest as the characterization of nanostructured materials becomes increasingly challenging to traditional methods. Moreover, scatterometry is far faster than full-field EUV CDI, which can give spatially and depth-resolved maps \cite{esashi2023tabletop,wang2023high,tanksalvala2021nondestructive} at the expense of extensive computation. Most notably, in addition to higher throughput, EUV scatterometry also results in reduced sample damage compared with electron microscopy \cite{wang2023high,jenkins2024euv} and excellent accuracy for both geometric and elemental properties over relevant sample areas \cite{esashi2023tabletop,herrero2021uncertainties}. Moreover, it often provides complementary information to other metrologies such as scanning electron or atomic force microscopy \cite{esashi2023tabletop,jenkins2024euv}. In the typical formulation, scatterometry of periodic structures yields information about the average unit cell in the sample, while imaging techniques produce a spatially resolved map of reflectance or optical density/transmission. However, for applications such as process control, information about large-scale variations across a printed wafer are of interest, and scatterometry has a significantly lower measurement time per unit area than full-field imaging.

Advances in HHG have made EUV scatterometry easily accessible in a robust tabletop setup \cite{kapteyn2023book}. In HHG, an intense ultrafast laser pulse is focused into a gas in a phase-matched geometry, enabling the generation of bright coherent beams spanning the EUV spectral region. These capabilities are enabling accurate EUV metrologies that can support the rapid development of next-generation nanostructured materials and devices, as well as enabling fundamental studies into the functional properties of materials that determine device performance. Furthermore, one unique capability of HHG is a large degree of tunability of the harmonic comb across the EUV spectrum using multiple approaches \cite{bartels2000shaped,ryan2023optically}. EUV scatterometry measurements can benefit greatly from this tunability, and from the development of a comprehensive approach to optimal experimental design \cite{klein2025spie}.

In this paper, we use a 29 nm HHG source to perform EUV scatterometry on single-nanometer scale out-of-plane features on a sample consisting of ortho-linear 2D periodic structures that are representative of a typical hybrid bonding interface. In a hybrid bonding application, two interfaces with the same structures are attached so that the dielectric area and metal areas are aligned. To achieve bonding the sample must be heated which also thermally expands the metal area. Therefore, defects such as voids or broken bonding can occur due to the lack or excess size of the dishing depth. Thus, precise topography measurements are a critical metrology challenge for semiconductor devices. Existing scanning electron microscopy fails to capture three-dimensional features without stereographic methods, especially when sample composition varies along with topography. Moreover, atomic force microscopy is limited by low throughput and can only measure small areas. Here we show how EUV scatterometry can measure both the nanoscale 3D structure of the sample as well as the materials that compose it by fitting a forward model to the measured diffraction efficiency data using rigorous coupled wave analysis (RCWA). This provides an abundance of information about the sample, which could inform the polishing and lithographic procedures that produced it. In the future, by reducing sample contamination due to sample preparation, deep sub-nanometer sensitivity to axial profiles and 2D structures will be possible on arbitrary samples, for example to provide feedback and efficient process control with high throughput.

Finally, we show how EUV scatterometry measurements can be optimized by taking advantage of the wavelength tunability and additional properties (e.g. polarization and multiple harmonics) of HHG sources as well as the ability to select the incidence angles. Traditional sensitivity analysis tools based on the Fisher information matrix (FIM) are well suited for this purpose and have been applied in other areas of research for optimizing experimental design \cite{klein2025spie,durant2022optimizing,mikhalychev2021fisher,yang2024decision,weedon1991selection,durant2021determining,dong2014determination}. Thus, by combining an analytic noise model with the simulated diffraction efficiency, we construct the FIM and perform computationally tractable optimizations of the normalized FIM eigenvalues to maximize the sensitivity of EUV scatterometry measurements. We perform the optimizations assuming two important unknown sample parameters and leave the incidence angles and wavelength as free experimental design parameters. The optimizations show that a wavelength just above the silicon absorption edge at $\sim$14 nm, well within the range of tunability of HHG, will maximize the sensitivity of EUV scatterometry measurements for the sample considered. This is due primarily to the greater penetration depth of the shorter wavelength EUV light. Moreover, correlations among model parameters have been identified as a key source of modeling error in scatterometry in the past \cite{herrero2021uncertainties,orji2018metrology,ansuinelli2019automatic}; in this regard, we show that the incidence angles are of great importance and must be carefully selected to ensure the data points are uncorrelated and with high-sensitivity to permit an accurate reconstruction of the sample, which the methods used here can optimize for. Given the strong dependence of measurement sensitivity on experimental design parameters even in this simple two-parameter case, a methodology for optimizing these parameters is an essential part of accurate quantitative EUV scatterometry with high throughput for general samples.

\section{Experimental Setup}

\begin{figure}[htbp]
\centering\includegraphics[width=4.5in]{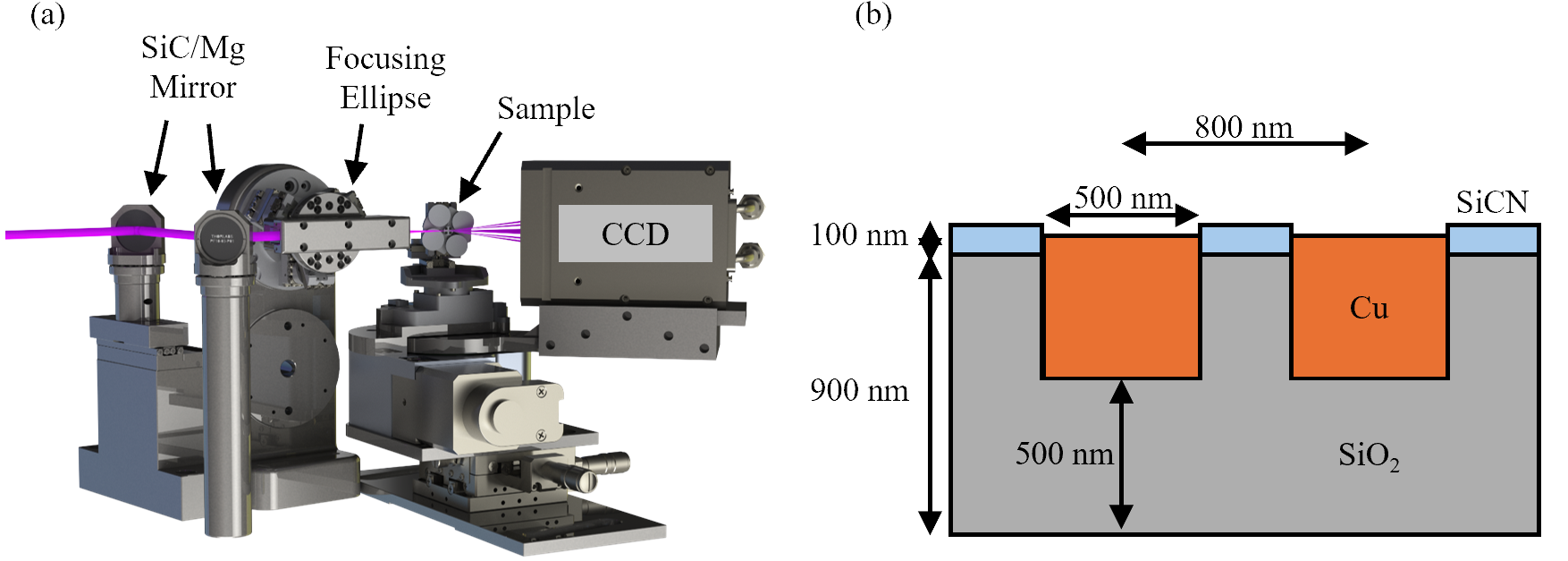}
\caption{(a) Schematic diagram of the experimental chamber, consisting of two multilayer mirrors, a focusing ellipse, the sample, and a CCD camera all under vacuum. (b) Schematic diagram of two unit cells of the sample as viewed from the side.}
\label{fig:1}
\end{figure}

Our experiment uses a titanium sapphire regenerative laser amplifier system (KMLabs Inc.) with $\sim$2.6 mJ pulse energy, $\lambda=780$ nm, $\tau_{pulse}\sim35$ fs, and 3 kHz repetition-rate to produce high harmonics in argon gas. The resulting harmonic comb is then sent through a set of two silicon rejector mirrors and aluminum filters that eliminate the infrared driving light. Two SiC/Mg multilayer mirrors (NTT-AT) designed to have a peak reflectance at 29.4 nm are then used to produce quasi-monochromatic EUV light. The EUV beam is focused with an ellipsoidal mirror to a spot size of $\sim$2 $\mu$m on the sample with s-polarization. The sample and an in-vacuum CCD camera are mounted in a $\theta$-$2\theta$ configuration so that the diffracted wavefront is recorded at a series of incidence angles. A schematic diagram of the experimental setup is shown in Fig. \ref{fig:1}(a). Further explanation of the experimental setup and its capabilities are described in Esashi \textit{et al}. \cite{esashi2023tabletop}, while ref. \cite{kapteyn2023book} describes near state-of-the-art HHG systems.

The sample we measured is a 2D ortho-linear periodic interconnect structure (Samsung) consisting of an 800 nm period array of square copper pads 500 nm wide that are embedded in a substrate, composition as shown in Fig. \ref{fig:1}(b). A key feature of this sample is that the copper pads are recessed into the sample by a small amount (a few nanometers) as a result of CMP planarization. The ability to measure this dishing depth with high throughput is a critical capability for the creation of a functioning device, as a precise value is needed to form good contacts between adjacent device layers in subsequent fabrication steps. Thus, the primary parameter we are interested in measuring is the average dishing depth between the copper pads and the surrounding substrate. However, due to uncertainties in other sample properties, such as the density of the SiC$_x$N$_y$, we also simultaneously calculate other quantities by including them as free parameters in the reconstruction algorithm. A schematic diagram of the sample is shown in Fig. \ref{fig:1}(b).

\section{Experimental Results}

\begin{figure}[htbp]
\centering\includegraphics[width=4.5in]{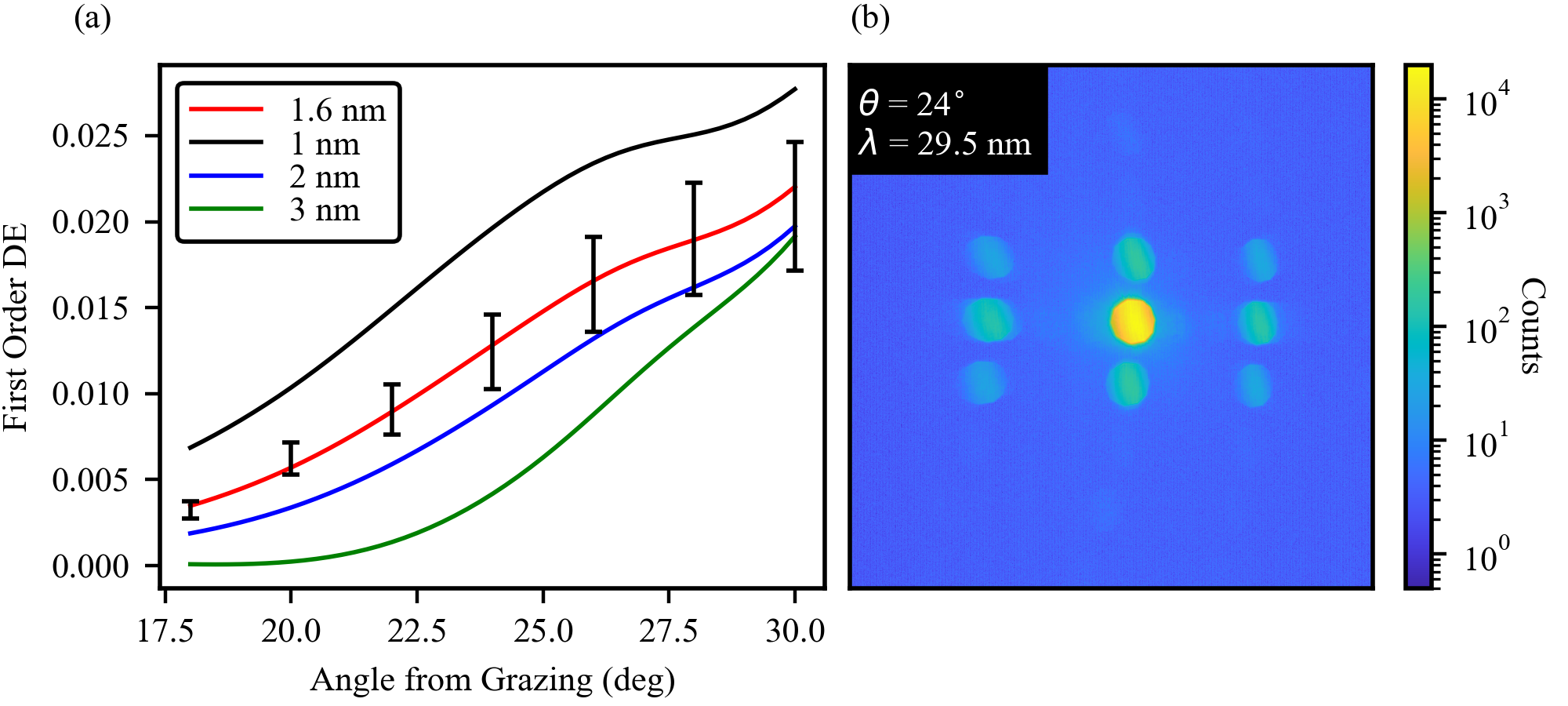}
\caption{(a) Experimental data shown as error bars indicating the standard deviation of a single measurement, and the model fit to the data (in red) along with diffraction efficiency curves resulting from single nanometer variations in the average dishing depth. Note that the experimental error bars are higher than expected on this sample due to the presence of debris after sample dicing that could not be cleaned. (b) Sample image showing the average of 50 diffraction images collected at a nominal incidence angle of 24$^\circ$ and wavelength of 29.5 nm.}
\label{fig:2}
\end{figure}

The diffraction patterns from the sample were recorded on an in-vacuum imaging CCD (PI-MTE2) at a series of seven incidence angles. The CCD integration time is set per-angle (ranging from .3-1.4 seconds per image) to nearly saturate the undiffracted (DC) order to optimize signal-to-noise. During CCD readout, the beam is blocked by an in-vacuum shutter. At each incidence angle, a 5x5 square array of probe positions with step sizes of 20 $\mu$m between positions is measured with 2 frames recorded at each position for a total of fifty diffraction images. Duplicate diffraction images were recorded to provide an estimate of the uncertainty of each angle measurement and average out noise and heterogeneities across the scan area. After background subtraction, for each diffraction frame, we select the positive and negative first-order conical-mount diffraction and DC order by identifying the region of the relevant pixels and compute the total counts by summing pixels within the region of interest. The first-order diffraction efficiency is then computed as
\begin{equation*}
    DE = \frac{N_{+1}+N_{-1}}{N_{+1}+N_{-1}+N_{0}},
\end{equation*}
for total CCD counts $N_n$ in the $n$th diffraction order, and the 50 diffraction efficiency values from the 50 diffraction images at each incidence angle are averaged. These repeated measurements reduce the measurement uncertainty since the standard deviation of the mean scales as the inverse of the square root of the number of diffraction efficiency values obtained from each image. The extracted average diffraction efficiency and their standard deviations are the data plotted in Fig. \ref{fig:2}(a), and an example of the diffraction images is shown in Fig. \ref{fig:2}(b).

The experimentally measured diffraction efficiency values were simulated using RCWA \cite{hugonin2021reticolo}. The model used in this simulation was then wrapped with a genetic algorithm \cite{ulyanenkov2000genetic} to provide a global optimization of the average chi-square error metric of the 7 sample points. Several sample parameters in the model were left as free parameters due to uncertainty in their exact value. These include the average copper dishing depth, SiCN density, and system calibration parameters of a global angle offset as well as a small offset in the EUV wavelength. Additionally, to account for unintentional contamination present on the sample from the cleaving/dicing process which could not be fully removed, a thin carbon layer of fixed thickness was added; this is an artifact that would not be present in an industrial application of scatterometry, and its effects do not change the general methodology we use here. Note that the experimental error bars are higher than expected due to the presence of this debris. The resulting fit to the data is shown in Fig. \ref{fig:2}(a), along with curves showing single nanometer variations to the average dishing depth and calculating a dishing depth of $1.60\pm.05$ nm.

\begin{figure}[htbp]
\centering\includegraphics[width=4.5in]{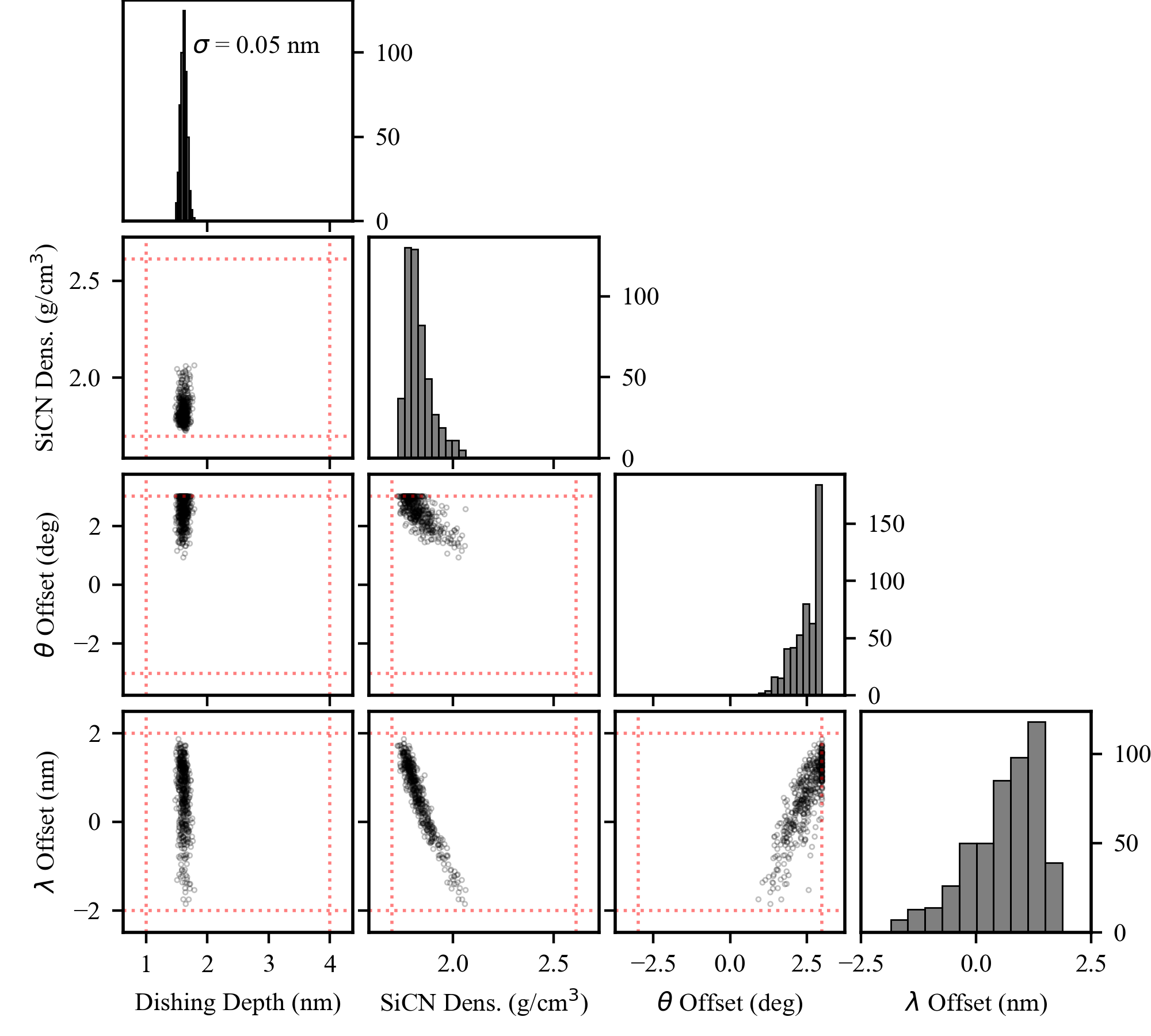}
\caption{Corner plot of the Monte Carlo results. Red lines indicate the priors that constrained the global optimization. Despite the correlations between the SiCN density, angle offset, and wavelength offset, an uncertainty of only .5 \r{A} is obtained for the dishing depth. These results can be further improved with the methods described in the following section.}
\label{fig:3}
\end{figure}

A critical aspect of metrology lies in quantifying the measurement uncertainty. Monte Carlo methods have proven to be an informative approach for uncertainty quantification in scatterometry \cite{soltwisch2017reconstructing,herrero2021uncertainties,gross2008computational,heidenreich2013alternative}. We performed a Monte Carlo simulation of the reconstruction algorithm where the data points used in each reconstruction are sampled from a Gaussian distribution with mean given by the average measured diffraction efficiency and standard deviation given by the standard deviation of the mean, $\sigma_{DE}/\sqrt{50}$, where 50 diffraction efficiency values are averaged at each incidence angle and $\sigma_{DE}$ is the standard deviation of the diffraction efficiency measurements from each of the 50 images, which provides an estimate of the uncertainty of a single frame. This is effectively the same as evaluating the uncertainty on model parameters that would be obtained if the experiment were repeated many times, since the mean value of the diffraction efficiency may change slightly for each experiment. 500 reconstructions, where each is an independent genetic algorithm optimization, were performed in this way to obtain an estimate of the standard deviation of the parameters that were measured; the full results of the Monte Carlo uncertainty evaluation are shown in Fig. \ref{fig:3}. We note that the uncertainty computed in this way will often be larger than when it is computed using the covariance matrix, which yields a theoretical lower bound on the parameter variances \cite{gross2008computational,heidenreich2013alternative}. To optimize these results and ensure the measurements are as accurate as possible, we propose the methods in the following section.

\section{Optimized Experimental Design}

Driven by broad bandwidth Ti:Sapphire lasers, HHG sources can be spectrally and polarization tuned throughout the EUV, with good flux and stability \cite{kapteyn2023book,ryan2023optically,fan2024efficient}. These capabilities can enhance the accuracy of EUV scatterometry. Thus, computationally tractable methods that can be used to guide experimental design before performing an experiment are an essential tool. Here, we show how optimization of parameters such as the EUV wavelength and incidence angle can be performed with the Fisher information matrix (FIM). The sample considered here is the same interconnect sample as in the previous section but with two parameters of interest – the average dishing depth and the incidence angle offset. This two-parameter case is a common case of interest in an industrial application where the average dishing depth is of primary concern, and it provides a clear demonstration of the methodology. Additionally, small uncertainties in angle offsets (of $\sim$1$^\circ$) are common even in commercial scientific and metrology instruments and have posed challenges for various forms of metrology in the past \cite{herrero2021uncertainties,klein2023computer}. The extension to more parameters is mathematically trivial, though computation times will increase as the number of optimization parameters increases and a global optimization of the eigenvalue becomes increasingly challenging since more local minima likely exist in the higher dimensional parameter space.

\begin{figure}[htbp]
\centering\includegraphics[width=4.5in]{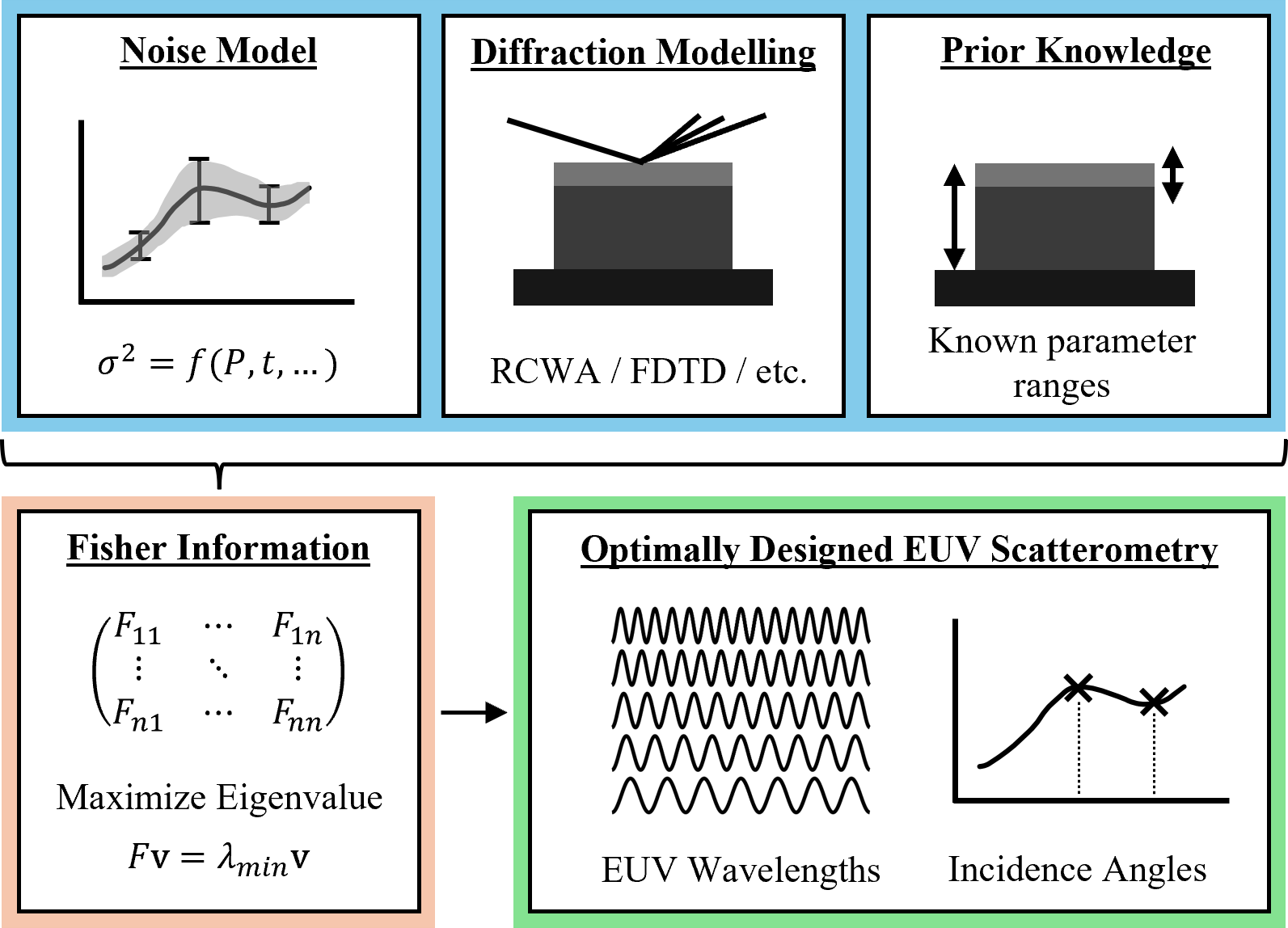}
\caption{A conceptual overview of the experimental optimization methodology. A rigorous noise model capturing fundamental physical uncertainties and experimental limitations is combined with diffraction modeling and prior sample knowledge to form the normalized Fisher information matrix. The minimum eigenvalue of the normalized Fisher information matrix is then maximized to predict the optimal experimental conditions. In this paper, we consider a two-parameter sample and determine the EUV wavelength and the incidence angles to sample the diffraction efficiency curve at that will maximize sensitivity. This methodology can easily be extended to accommodate additional/alternative experimental parameters such as polarization, sample face rotation, and additional EUV wavelengths which may provide an opportunity to increase sensitivity and throughput.}
\label{fig:4}
\end{figure}

Optimizing the experimental design of an EUV scatterometry experiment requires three key components: a noise model, diffraction modeling through a Maxwell’s equations solver such as RCWA, and prior knowledge about the sample. These three items are mathematically combined to form the normalized FIM, which can then be used to optimize experimental design parameters such as wavelength and incidence angles. A conceptual overview of the methodology we use in this paper is given in Fig. \ref{fig:4}.

We use a noise model that accounts for Poisson photon detection statistics as well as common experimental uncertainties introduced by the dark current and read noise of a typical CCD camera (see Supplemental Document). In the high flux regime, as in the case of typical EUV scatterometry experiments, we can regard the probability distribution of each diffraction efficiency measurement to be approximately Gaussian distributed \cite{park2006bayesian}. With this approximation, the variance of the diffraction efficiency takes the form
\begin{equation}\label{eq:1}
    \sigma_{DE}^2 = DE_m^2 \left[ \frac{\sigma_{N_{\pm m}}^2}{(N'_{\pm m})^2} + \frac{4\sigma_{N_{\pm m}}^2 + \sigma_{N_{0}}^2}{(2N'_{\pm m} + N'_0)^2} \right],
\end{equation}
where for the mth diffraction order, $DE_m$ is the diffraction efficiency, $\sigma_{N_{m}}^2$ is the variance on the total CCD counts with camera noise, and $N'_m$ is the total number of CCD counts after a background subtraction. We also set a hard upper limit of 5 minutes on the integration time required for the DC order to fill the full well capacity of the CCD camera to ensure reasonable experiment times and to prevent the accumulation of other stochastic sources of error such as cosmic rays and thermal fluctuations. If optimizing throughput is of interest, one may not need to fill the full well capacity of the detector to achieve the desired level of uncertainty on the model parameters and integration times can be reduced. The reader is invited to read the full details of the noise model in the Supplemental Document as well as additional considerations for throughput in the Discussion.

The matrix elements of the FIM are defined as
\begin{equation*}
    F_{kl} = \textrm{E}\left[ \left( \frac{\partial}{\partial p_k}\ln P(\mathbf{p}) \right) \left( \frac{\partial}{\partial p_l}\ln P(\mathbf{p}) \right) \right],
\end{equation*}
where $P(\mathbf{p})$ is the joint probability distribution of the diffraction efficiency at each angle for a set of model parameters $\mathbf{p}$. However, under the assumption of Gaussian distributed probability distributions for each measurement and using the fact that each measurement is independent, the FIM takes the simple form (see Supplemental Document)
\begin{equation*}
    F_{kl} = \sum_{i=1}^n \frac{1}{\sigma_i^2}\frac{\partial f_{\theta_i}}{\partial p_k}\frac{\partial f_{\theta_i}}{\partial p_l}.
\end{equation*}
In the case of a diffraction efficiency model, the $\sigma_i$ are the uncertainties on $f_{\theta_i}$ at each angle $\theta_i$ as given by Eq. \ref{eq:1}, and $f_{\theta_i}$ is the diffraction efficiency at $\theta_i$.

Since each model parameter has different units, the FIM must in general be normalized so that each parameter is on the same scale. Several approaches are taken in the literature for doing this, all of which scale the parameter values to some expected range based on the model uncertainty; we take the approach described by Durant \textit{et al}., which they applied to neutron reflectometry \cite{durant2022optimizing}. In this approach, each parameter $p_k$ is assigned some range over which the true parameter value is expected to lie $[a_k,b_k]$, and this parameter range is mapped to $[0,1]$. The transformation matrix for the two-parameter case is then $\mathbf{J}=\textrm{diag}[1/(b_1-a_1),1/(b_2-a_2)]$ and the FIM is transformed as $\mathbf{J}^T \mathbf{F} \mathbf{J}$.

With the FIM normalized, we can select a suitable optimization criterion that will maximize the sensitivity of EUV scatterometry. Many optimization criteria exist in the literature and an excellent overview of several is given by Mikhalychev \textit{et al}. \cite{mikhalychev2021fisher}. Since our reconstruction procedure with a genetic algorithm simultaneously varies all model parameters by random amounts, we must consider the sensitivity of the least sensitive direction of the system. Therefore, we take the approach of maximizing the smaller of the two eigenvalues. This maximizes the sensitivity of the system’s principal direction which is least sensitive to simultaneous parameter variations; additional details of this approach can be found in \cite{durant2022optimizing,yang2024decision}. In the past, other works for different types of metrology have minimized the Cramér-Rao bound for the parameters of interest \cite{weedon1991selection,durant2021determining}, however this may not necessarily give the best chance at an accurate reconstruction, and therefore maximizing the minimum eigenvalue is a preferable approach for the problem at hand.

\begin{figure}[htbp]
\centering\includegraphics[width=4.5in]{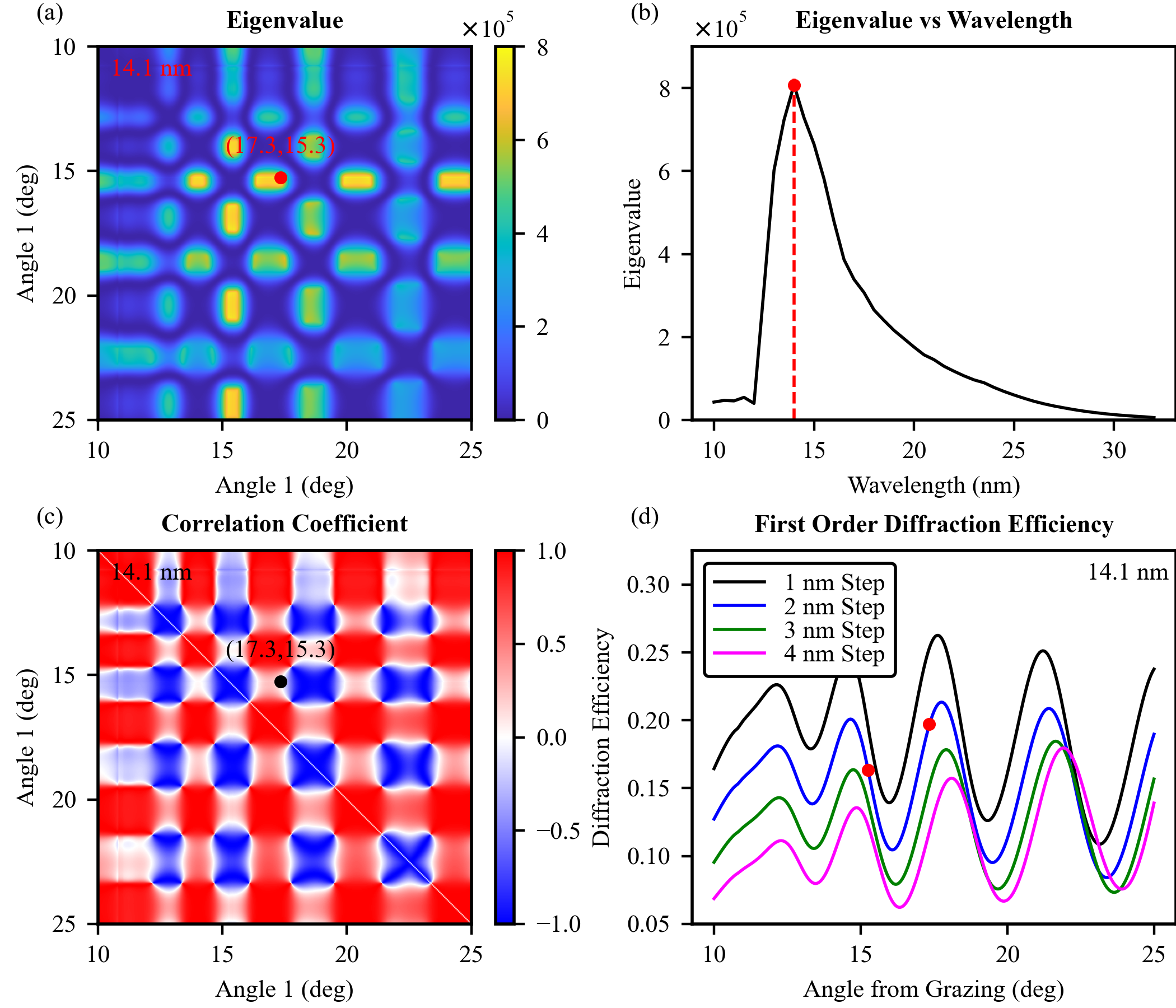}
\caption{(a) The minimum eigenvalue of the Fisher information matrix as a function of two incidence angles for the optimized wavelength of 14.1 nm. (b) Value of the minimum eigenvalue as a function of wavelength for the optimized incidence angles at each wavelength. (c) Correlation coefficient between the two model parameters as a function of the two incidence angles. (d) Sample diffraction efficiency curves at 14.1 nm for several dishing depths, and the optimized points (red dots) for a 2 nm dishing depth.}
\label{fig:5}
\end{figure}

A global optimization of the eigenvalue of the normalized FIM for the sample was performed with the illuminating wavelength and two incidence angles as free parameters. The optimization maximized the eigenvalue of the normalized FIM at a wavelength of 14.1 nm and incidence angles of 15.3$^\circ$ and 17.3$^\circ$. A plot of the eigenvalue as a function of incidence angle at this wavelength is shown in Fig. \ref{fig:5}(a), and the eigenvalue at the optimal angles for each wavelength as a function of wavelength is shown in Fig. \ref{fig:5}(b). Additionally, a plot of the correlation coefficient as a function of the incidence angle and several sample diffraction efficiency curves at 14.1 nm are shown in Fig. \ref{fig:5}(c) and (d). The maximization of the eigenvalue identifies the optimal incidence angles and wavelength by linking the models of the noise and diffraction efficiency as well as inter-parameter correlations together in a mathematically rigorous model.

\section{Discussion}

The results in Figs. \ref{fig:5}(a) and (b) show that both the incidence angle and the illuminating wavelength significantly affect the sensitivity of EUV scatterometry even for this simple two-parameter sample. As more parameters are introduced, the interplay between each model parameter on the features of the diffraction efficiency curves will become more complicated, and an intelligent selection of the incidence angles through this methodology will be increasingly important. Consequently, thoughtful experimental design, unique to the specific sample/quantity of interest, is a critical component to the success of accurate EUV metrology.

The unique features of the specific case presented in Fig. \ref{fig:5}(a) can be understood in a simple manner. As is evident in Fig. \ref{fig:5}(d), for this particular sample the effect of increasing the dishing depth is to nearly everywhere reduce the diffraction efficiency (i.e. the derivative of the diffraction efficiency with respect to dishing depth is negative for nearly all incidence angles). On the other hand, the effect of increasing the angle offset (effectively shifting the curves left) either increases or decreases the diffraction efficiency depending on the sign of the derivative with respect to incidence angle of the diffraction efficiency curve at each point. Therefore, when one incidence angle lies in a region of positive derivative with respect to angle and the other angle lies in a region of negative derivative with respect to angle, as is the case for the two optimized points in Fig. \ref{fig:5}(d), the correlation between the dishing depth and angle offset will be broken and this will tend to increase the eigenvalue of the normalized FIM, as can be seen by comparing Figs. \ref{fig:5}(a) and (c). The fact that the two optimized red points in Fig. \ref{fig:5}(d) lie on those specific oscillations as compared to any other oscillation is primarily determined by the noise model, which tends to reduce the eigenvalue at larger angles of incidence due to the accumulation of noise as a result of the larger integration times from decreasing sample reflectivity. Furthermore, since any oscillatory diffraction efficiency curve with respect to incidence angle will necessarily have a derivative with respect to angle offset which oscillates between positive and negative, this strong dependence of the measurement sensitivity on incidence angle will be a universal feature of samples with oscillatory diffraction efficiency curves. Interference effects from multilayer and near-wavelength-sized transverse structures are the main contributors to oscillatory diffraction efficiency vs. incidence angle curves and will therefore be nearly ubiquitous features in diffraction metrology at wavelengths of $\sim$14 nm.

\begin{figure}[htbp]
\centering\includegraphics[width=5in]{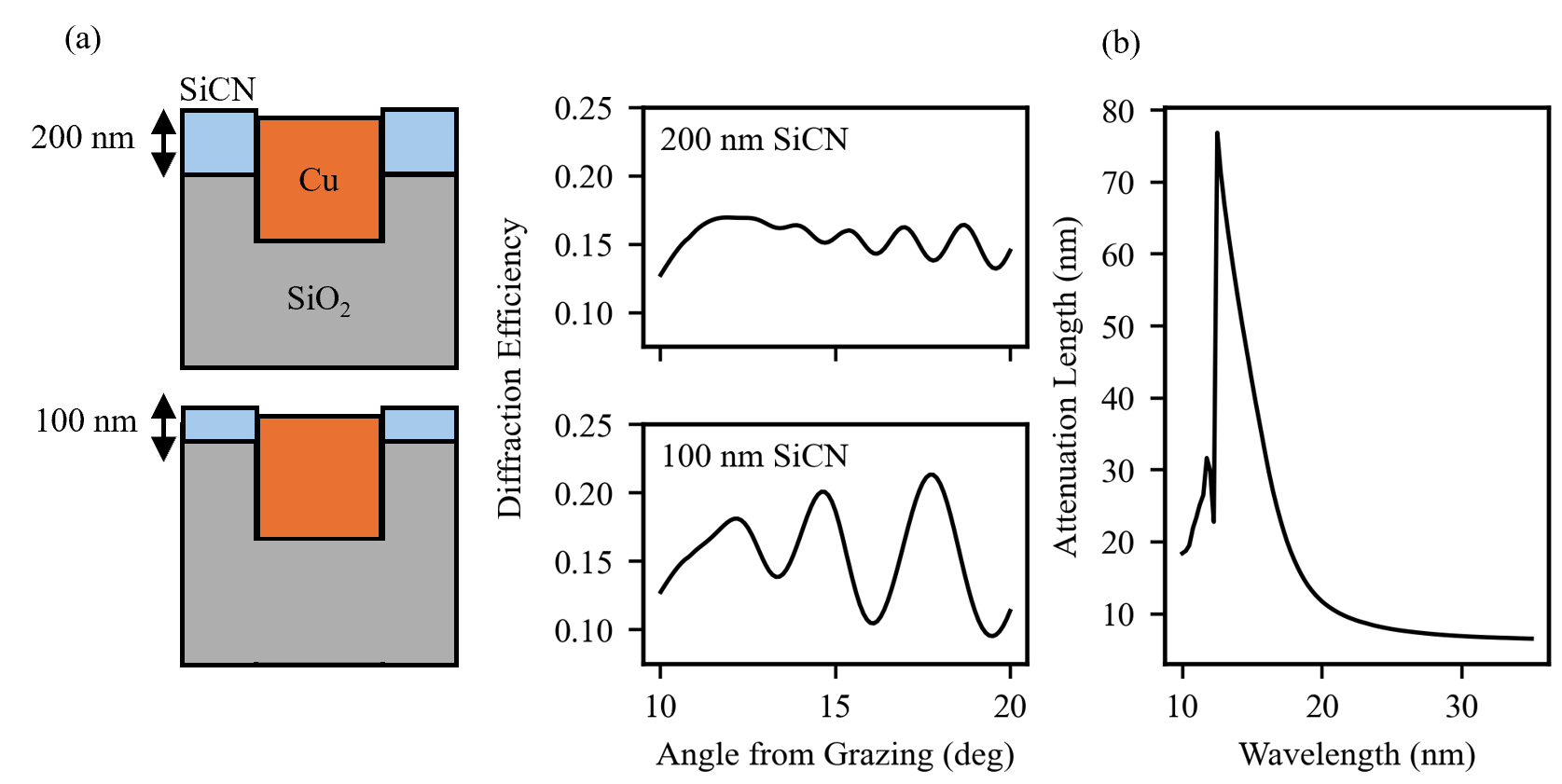}
\caption{(a) The higher penetration depth of 14 nm EUV light and the thin SiCN layer leads to the oscillatory behavior in the diffraction efficiency curves, which enables a more sensitive measurement of the sample considered for reasons explained in the Discussion. As the thickness of the SiCN layer increases the oscillatory behavior is lessened. (b) Attenuation length of EUV light as a function of wavelength in the SiCN layer for a 15$^\circ$ angle of incidence from grazing. Note the strong similarity between the shape of the attenuation length curve and the eigenvalue in Fig. \ref{fig:5}(b), indicating this feature of the EUV light is a key factor that enables a more sensitive measurement. Data from the CXRO database \cite{henke1993x}.}
\label{fig:6}
\end{figure}

The optimized wavelength can be understood as a consequence of the greater penetration depth of the shorter wavelength EUV light as demonstrated in Fig. \ref{fig:6}. Since the sample consists of a 100 nm surface layer of SiCN, the shorter wavelength EUV light at $\sim$14 nm is able to penetrate this layer and produce the interference/oscillatory effects that are useful for the reasons described in the previous paragraph. We note that even though we are only interested in a surface feature (the dishing depth) with an additional calibration parameter of the angle offset, the greater penetration depth of 14 nm light proves to be useful. It should also be noted that the prior knowledge of the sample as captured in the FIM formalism affects the value of the optimized wavelength and is an important part of the calculation; if one can set the incidence angle very precisely so that the uncertainty is reduced, this will rescale the matrix elements of the normalized FIM and reduce the relative importance of the interference fringes on the measurement sensitivity, thus shifting the optimized wavelength to slightly larger values. 

We emphasize that while the overall intuition behind the angle regions and wavelength where measurement sensitivity is high can be easily understood for this two-parameter case, an intuitive understanding of the case with more parameters is not as straightforward. However, the methodology using the FIM is easily extended to more parameters. In general, for a single wavelength the number of incidence angles required will be equal to the number of unknown model parameters, and the incidence angles must be optimized to simultaneously break correlations between all model parameters while also considering the noise model and the dependence on wavelength. The FIM is therefore an essential tool for guiding experimental design to ensure that the measurements resulting from EUV scatterometry reconstructions are accurate and reliable with minimized variance on parameters of interest.

An important aspect that influences the optimized experimental parameters is the noise model that is used. To obtain experimentally practical results from the FIM methodology, one must consider the sources of noise in the experiment and the limitations of experimental equipment. The presence of camera noise in the model is an important feature since it accurately captures the low sensitivity of near-grazing and near-normal incidence angles whereas a model of, for example, only photon noise may converge to experimentally impractical incidence angles. If one would like to place additional limitations on other aspects of the experiment, such as the integration time at each incidence angle to ensure a high throughput, these can also be captured by the noise model, although having such limitations may impose a reduction on the sensitivity. The methods used in this paper can also be adapted to other forms of measurement such as reflectometry and imaging reflectometry provided that the noise model is altered for the measured signal in those cases.

A particularly interesting future application of this experimental design methodology is to constrain the dose applied to a sample. As HHG is driven by more efficient long wavelength mid IR lasers \cite{morrill2025soft}, harmonics can be extended to soft x-ray (SXR) wavelengths which span the water window (2.3-4.4 nm). With bright tabletop SXR sources in the water window, a wider variety of soft material samples enter the purview of CDI quantitative microscopy. However, one must take care to measure biological or polymer samples with an appropriate dose of energy (J/cm$^2$) to avoid damaging them. This consideration is especially important for hydrated biological samples, which can undergo many damage mechanisms \cite{jacobsen2019microscopy,du2018relative} that reduce the possible spatial resolution by limiting the measured diffraction signal. Alternative CDI schemes such as in-situ CDI \cite{lo2018situ,lu2024computational} can enable lower dose and possibly dynamic imaging of hydrated samples through coherent interference with a neighboring static structure. When quantitative analysis is the goal, the optimality of measurement conditions should also be considered. Practically, our methodology provides a pathway to address such a constraint, with the most straightforward application being a selection of the illuminating wavelength given a dose limitation on the sample. Similar to scatterometry, in the case of CDI one must consider the noise model of the imaging system, dose limitations of the sample, and prior knowledge about sample geometry and stoichiometry.

An additional consideration for any practical measurement/experiment using this methodology to optimize incidence angles, wavelength, and any other parameter, is that there is inherently some uncertainty about the model parameters prior to the measurement. In the case presented here, it is evident that for the optimized points in Fig. \ref{fig:5}(d), the oscillatory regions where the derivative is positive or negative remain at about the same location for the range of uncertainty on the dishing depth. However, the uncertainty due to a global incidence angle offset may present a challenge. In general, more uncertainty may require alternative data, and if one wishes to fully maximize the sensitivity of the measurement, it may be necessary to allocate the integration time across a larger number of incidence angles. This can account for the uncertainty in the model, and by collecting data across this range during the one initial experiment, the key points where sensitivity is maximized can be targeted.

For increasing throughput, the topic of interest is the experimental conditions that will achieve the desired level of measurement uncertainty with the smallest amount of measurement time possible; if one is aiming to achieve a specified level of uncertainty on model parameters with high throughput, then one should not integrate the signal any longer than is required to achieve that uncertainty. Regarding this, the methodology used in this paper can help to increase throughput by obtaining the most sensitive measurement conditions, which will reduce the integration time and amount of data needed to achieve the desired uncertainty. Suboptimal measurement conditions will lead to suboptimal throughput; therefore, one must ascertain that measurement time is being spent on measuring a useful signal and not on additional data that may not provide a strong contribution to sensitivity. Additionally, with careful consideration of the experimental capabilities one has access to, various multiplexing schemes can be considered to increase throughput further; these may include using additional wavelengths in a harmonic comb or fitting additional diffraction orders that may rise above the noise floor at certain wavelengths. With the methods used in this paper, throughput can be optimized to help overcome e.g. flux limitations as well as reducing the total dose incident on the sample.

\section{Conclusion}

We performed an experimental demonstration of EUV scatterometry on a 2D periodic interconnect sample, calculating an axial dishing depth of $1.60 \pm .05$ nm. Then, through an experimental optimization procedure based on the Fisher information matrix, we showed that both the choice of incidence angle and the specific EUV wavelength used can have a significant effect on the sensitivity of measurements made with EUV scatterometry. We also described how the strong dependence of sensitivity on incidence angle is a general feature of oscillatory diffraction efficiency curves. The experimental optimization method presented is therefore an essential component of EUV scatterometry to ensure accurate measurements, and the methods can be readily extended to provide an intelligent design of experiments that enable full use of the tunable properties of HHG-based EUV light, such as wavelength, polarization, and additional harmonics, as well as the general orientation of the incident light relative to the sample, with optimized throughput.

\begin{backmatter}

\bmsection{Acknowledgments}
This research was performed at JILA, University of Colorado. The authors gratefully acknowledge support from the STROBE NSF Science and Technology Center, Grant No. DMR-1548924. We also acknowledge support from Grant No. AWD-22-06-0106 from Samsung Telecommunications America, LLC, and Samsung for providing the sample. We acknowledge U.S. Air Force Office Multidisciplinary University Research Initiative (MURI) program under award no. 487 FA9550-23-1-0281 for the low-dose imaging applications.

\vspace{-2.5pt}
\bmsection{Disclosures}
MMM: Kapteyn-Murnane Laboratories (I,S), HCK: Kapteyn-Murnane Laboratories (F,I,E,S).

\vspace{-2.5pt}
\bmsection{Data Availability}
The data from this paper are not publicly available at this time but may be obtained from the authors upon reasonable request.

\vspace{-2.5pt}
\bmsection{Supplemental document}
See Supplement 1 for supporting content.

\end{backmatter}

\vspace{-2.5pt}
\bibliography{ms}

\end{document}


\maketitle

\section{Noise Model for the Diffraction Efficiency}

For a typical CCD camera, the number of counts recorded per pixel is
\begin{equation*}
    N_{pix} = P_n Q_e t + Dt + N_r
\end{equation*} 
where $P_n$ is the number of photons incident on the pixel $n$ per unit time, $Q_e$ is the quantum efficiency, $t$ is the integration time, $D$ is the dark current, and $N_r$ is the read noise. To ultimately obtain an analytic equation for the signal to noise of the diffraction efficiency and thus enable efficient optimizations of experimental design, we make the approximation that each diffracted order is evenly spread over $N$ pixels so that for a total flux $P$ incident on the sample the total number of counts in the $m$th diffraction order is given by
\begin{equation*}
    N_m = P \phi_m Q_e t + DtN + N_rN.
\end{equation*}
where $\phi_m$ is the fraction of the flux incident on the sample that gets diffracted into the $m$th order. The assumption that the spot size is the same for different wavelengths is reflected in the fact that the EUV beam size can be experimentally changed with an iris prior to its diffraction from the sample.

In a typical scatterometry experiment, we set the integration time so that the DC order fills some fraction $\alpha$ of the full well capacity $C$ of the camera so that the integration time satisfies $P \phi_0 Q_e t = \alpha C N$. Thus the number of counts in the $m$th order is
\begin{equation} \label{m_counts}
    N_m = \frac{\phi_m}{\phi_0}\alpha CN + \frac{\alpha DCN^2}{P\phi_0 Q_e} + N_r N.
\end{equation}
The first two terms in Eq. \ref{m_counts} from the photon count and dark current are Poisson distributed, while the last term from read noise is normally distributed. Thus the variance of the CCD counts of a diffracted order is
\begin{equation} \label{m_error}
    \sigma_{N_m}^2 = \frac{\phi_m}{\phi_0}\alpha CN + \frac{\alpha DCN^2}{P\phi_0 Q_e} + N_r^2 N.
\end{equation}
We note that this is also the variance on the number of counts recorded after a background subtraction,
\begin{equation} \label{counts_bg}
    N_m' = \frac{\phi_m}{\phi_0}\alpha CN.
\end{equation}

The measured diffraction efficiency is defined to be
\begin{equation} \label{de}
    DE_m = \frac{N_{+m}'+N_{-m}'}{N_{+m}'+N_{-m}'+N_0'} = \frac{2N_{\pm m}'}{2N_{\pm m}'+N_0'},
\end{equation}
where the second equality results for the non-conical direction where $N_m=N_{-m}=N_{\pm m}.$ Eq. \ref{de} is a ratio of Poisson distributions, so the error cannot be propagated analytically. However, as we are in a regime of very high flux, we can follow the classical approach described by Park \textit{et al.}, which is a good approximation for high flux \cite{park2006bayesian}. In this approach we use Gaussian error propagation on the Poisson uncertainties, which yields
\begin{equation} \label{de_uncertainty}
    \sigma_{DE_m}^2 = DE_m^2 \left[ \frac{\sigma_{N_{\pm m}}^2}{(N_{\pm m}')^2} + \frac{4\sigma_{N_{\pm m}}^2 + \sigma_{N_0}^2}{(2N_{\pm m}'+N_0')^2} \right],
\end{equation}
where the variances are given by Eq. \ref{m_error} and the $N_m'$ are given by Eq. \ref{counts_bg}. 

The quantity $\phi_m$ is computed with a rigorous coupled wave analysis simulation and thus the variances $\sigma_{DE_m}^2$ can be computed for known camera noise parameters available in the manufacturer specifications. The values of the quantities used in the simulations in the main paper and their physical meaning are given in Table \ref{table1}. Additionally, to keep experimental throughput high and prevent the accumulation of other stochastic sources of experimental noise such as cosmic rays and thermal fluctuations, a hard upper limit of 5 minutes was set on the integration time. The inclusion of camera noise is an essential feature for experimentally practical results since the noise floor prevents the variance from approaching zero as the diffraction efficiency approaches zero for near-grazing angles, as would happen for a model which only considers Poisson noise on the photon count. It is straightforward to generalize this model to the case when $N_{+m} \neq N_{-m}$, which is of interest when using other sample rotation angles (e.g. face rotation) in experimental optimizations.

\begin{table}
\begin{center}
\begin{tabular}{ |c|c|c|c| } 
 \hline
 Parameter & Value & Units & Meaning \\ 
 \hline
 $P$ & $10^8$ & photons/sec & Total flux incident on sample \\ 
 $\alpha$ & .9 & NA & Fraction of full well capacity filled \\ 
 $Q_e$ & .9 & electrons/photon & Quantum efficiency \\
 $C$ & 150000 & electrons/pixel & Full well capacity \\
 $N$ & $50^2$ & pixels & Pixels per diffracted order \\
 $D$ & .0015 & electrons/pixel/second & Dark current \\
 $N_r$ & 6.5 & electrons/pixel & Read noise \\
 \hline
\end{tabular}
\caption{Camera noise parameters used in the simulations in the main paper.}
\label{table1}
\end{center}
\end{table}

\section{Fisher Information Matrix for Gaussian Error}
The Fisher information matrix simplifies to a well known form for Gaussian error and independent data points. We briefly repeat the derivation here to be clear about the underlying assumptions as it relates to scatterometry.

Suppose we have a function $f_{\theta_i}(\textbf{p})$ that describes the value of the diffraction efficiency at the angle $\theta_i$ for a vector of known model parameters $\textbf{p}$. The value of the random variable $y_i$ at each data point $\theta_i$ has a mean value of $f_{\theta_i}(\textbf{p})$ and follows a Gaussian distribution with standard deviation $\sigma_i$ given by Eq. \ref{de_uncertainty}. Thus, each observable has a probability distribution

\begin{equation*}
    P_{\theta_i}(\textbf{p}) = \frac{1}{\sqrt{2\pi\sigma^2}}\textrm{exp}\left[ -\frac{(y_i-f_{\theta_i}(\textbf{p}))^2}{2\sigma_i^2} \right],
\end{equation*}
Since each diffraction efficiency measurement is independent, the joint probability density function is
\begin{equation*}
    P(\textbf{p}) = \prod_{i=1}^{n}P_{\theta_i}(\textbf{p}).
\end{equation*}

The matrix elements of the Fisher information matrix (FIM) for this probability distribution and model parameters are then defined as
\begin{equation}\label{FIM}
    F_{ij} = \textrm{E}\left[ \left( \frac{\partial}{\partial p_i}\ln P(\textbf{p}) \right) \left( \frac{\partial}{\partial p_j}\ln P(\textbf{p}) \right) \right].
\end{equation}

We first evaluate the log-likelihood function. Using properties of the natural logarithm, we have
\begin{align*}
    \ln P(\textbf{p}) &= \ln \left[ \prod_{i=1}^{n}P_{\theta_i}(\textbf{p}) \right] \\
    &= -\frac{n}{2}\ln(2\pi\sigma^2) + \sum_{i=1}^{n} \left[ -\frac{(y_i-f_{\theta_i}(\textbf{p}))^2}{2\sigma_i^2} \right].
\end{align*}
Evaluating the derivative
\begin{align*}
    \frac{\partial}{\partial p_k} \ln P(\textbf{p}) &= \sum_{i=1}^{n} \frac{\partial}{\partial p_i} \left[ -\frac{(y_i-f_{\theta_i}(\textbf{p}))^2}{2\sigma_i} \right] \\
    &= \sum_{i=1}^{n} \frac{1}{\sigma_i^2} (y_i-f_{\theta_i}(\textbf{p}))\frac{\partial f_{\theta_i}}{\partial p_k}.
\end{align*}

With this result, we can now evaluate the matrix elements of the FIM using the definition Eq. \ref{FIM}:
\begin{align*}
    F_{kl} &= \textrm{E} \left[ \left( \sum_{i=1}^{n} \frac{1}{\sigma_i^2} (y_i-f_{\theta_i}(\textbf{p}))\frac{\partial f_{\theta_i}}{\partial p_k} \right) \left( \sum_{i=1}^{n} \frac{1}{\sigma_i^2} (y_i-f_{\theta_i}(\textbf{p}))\frac{\partial f_{\theta_i}}{\partial p_l} \right) \right] \\
    &= \sum_{i=1}^{n} \sum_{j=1}^{n} \frac{1}{\sigma_i^2\sigma_j^2} \frac{\partial f_{\theta_i}}{\partial p_k}  \frac{\partial f_{\theta_j}}{\partial p_l} \textrm{E} \left[ (y_i-f_{\theta_i}(\textbf{p}))(y_j-f_{\theta_j}(\textbf{p})) \right]. \label{fisher_intermediate} \numberthis
\end{align*}
Now using the fact that
\begin{equation*}
    \textrm{E} \left[ (y_i-f_{\theta_i}(\textbf{p}))(y_j-f_{\theta_j}(\textbf{p})) \right] = 
    \begin{cases}
        \sigma_i^2, & \textrm{if $i=j$}\\
        0, & \textrm{if $i\neq j$},
    \end{cases}
\end{equation*}
the double sum in Eq. \ref{fisher_intermediate} collapses and we are left with
\begin{equation*}
    F_{kl} = \sum_{i=1}^n \frac{1}{\sigma_i^2} \frac{\partial f_{\theta_i}}{\partial p_k}  \frac{\partial f_{\theta_i}}{\partial p_l},
\end{equation*}
which is the typical form for the matrix elements of the FIM with Gaussian error on each observable that we used in the main paper.

Since each model parameter $p_k$ in general has different units and are not on a comparable scale, the FIM must be normalized. In the main paper, this was done using the approach described by Durant \textit{et al}., which assumes the parameter values are known to lie within some range $r_k$ \cite{durant2022optimizing}. The FIM is then normalized so that
\begin{equation*}
    F' = J^T F J,
\end{equation*}
where $J = \textrm{diag}( 1/r_1, \dots 1/r_n )$ and $F'$ is the normalized FIM. For the two-parameter model we considered in the paper, we chose the range of the step height to be 2 nm and the range of the angle offset to be 5 degrees. Since our reconstruction procedure varies all model parameters simultaneously, the minimum eigenvalue was used as the optimization criterion, as is a typical approach \cite{yang2024decision,durant2022optimizing}.

\bibliography{supplement}